\documentstyle[aps,preprint,epsfig]{revtex}
\begin{document}
\title{\bf Calculating the Curie Temperature  reliably in diluted
III-V ferromagnetic semiconductors}

\author{Georges ~Bouzerar$^{1}$\footnote{email:bouzerar@ill.fr}, Timothy Ziman$^{1}$\footnote{and CNRS, email:ziman@ill.fr} and Josef ~Kudrnovsk\'y$^{2}$\footnote{email:kudrnov@fzu.cz}}

\address{ 
$^{1}$Institut Laue Langevin 
BP 156
38042 Grenoble 
France.\\            
$^{2}$ Institute of Physics, Academy of Sciences of the Czech Republic,Na Slovance 2,CZ-182 21 Prague 8, Czech Republic }
\date{\today}

\maketitle

\begin{abstract}
\parbox{14cm}{\rm}
\medskip
We present a semi-analytic theory for the Curie temperature in  diluted magnetic
semi-conductors that treats disorder effects exactly in the effective Heisenberg Hamiltonian, and spin fluctuations
within a local RPA. The exchange couplings are taken from  concentration dependent{\it ab initio} estimates.
The theory  gives very good agreement
with  published data for well-annealed samples of  
 Mn$_x$Ga$_{1-x}$As. We predict the critical temperatures for
 Mn$_x$Ga$_{1-x}$N  lower than in doped GaAs, despite
the stronger nearest-neighbour  ferromagnetic coupling. We also predict the dependence
on the hole concentration.

\end{abstract} 
Search for diluted magnetic semiconductors with ferromagnetism stable to high temperatures has been hampered by the many physical parameters that
may determine magnetic properties.
These include the choice of host semiconductor, that of the doping magnetic
impurity, the degree of compensation, and methods of preparation
and treatment of the sample\cite{Ohno}. 
The underlying mechanism of interaction between dopant
spins has without doubt been correctly identified:
Ruderman-Kittel-Kasuya-Yosida(RKKY)-like effective interactions
mediated by both the host\cite{Dietl,Jungwirth} and the doping band \cite{Akai,Josephetal,Georges3}.
Despite this, 
the theory has not lead to reliable quantitative predictions.
Comparison of calculations has been complicated 
by difficulties of full characterizing the samples.
In samples of GaAs doped with Mn, extensive  experimental studies\cite{Ohno}
have now allowed for greater control over sample parameters
and there is now apparently convergence to reliable experimental
values of the critical temperature of different groups\cite{Edmonds2,Edmonds,Matsukura,Chiba}. An important factor was that the carrier densities
were measured simultaneously by  
magneto-transport.
There is now the possibility of testing calculations against
experiments and determining the origin of past discrepancies.
\par
{\it Ab initio} calculations
using the Local Density Approximation and the magnetic force theorem
can be used\cite{Josephetal} to derive realistic
values of 
magnetic exchange interactions between classical impurity spins. These calculations
also take into account the effect on the effective exchange of 
disorder of the carriers, within a Coherent Potential  Approximation (CPA).
Similar calculations based on supercells\cite{vanSchilf,Sandratskii} lead to comparable results
at low concentration.
It has become apparent that the difficulty
is not in deriving the effective magnetic Hamiltonian correctly
but in 
treating its thermodynamics
accurately. As  we will demonstrate explicitly here, treating the magnetic correlations 
by over-simplified mean field theories\cite{Satoetal,Bhatt} has lead to overstatement , by a wide margin, of the critical temperature T$_c$.
The disorder in the effective
magnetic model also plays an important
role that cannot be simply treated by an effective medium theory
of the style of the Virtual Crystal Approximation (VCA)\cite{Georges2}.
This  suggests that the discrepancy with 
experiment is largely due to approximations made to the effective
Hamiltonian, not the values of the couplings themselves.
Thus by improving the treatment of the effective Hamiltonian we can hope
to find good agreement with experiments.
\par
In this paper we shall argue that
the effective random Heisenberg model may be treated 
by an accurate semi-analytic method separating the
treatment of the disorder, which will be treated without approximation,
and an analytic approach, a form of the Random Phase Approximation (RPA) for spin fluctuations.
Thus the calculation starting from first principles 
is in three stages: first the {\it ab initio} calculation of the effective Heisenberg couplings
with pairs of magnetic dopants at different
relative displacements.
In the second stage, 
we  generate a sequence of different  configurations 
on the fcc lattice by  sampling
techniques.
For each configuration the random Heisenberg model
is treated analytically within   RPA.
This approximation is an extension 
of standard RPA of the Heisenberg model  to a disordered system.
It 
is equivalent to that used in ref\cite{Georges2},
except  in that case the disorder was treated in a CPA-type
manner.
As the lattice
configuration is random, the equations must be solved numerically.
The full derivation of the equations
will be given elsewhere\cite{Bouzerarmethod}: here we shall simply summarize
the  determination of the the critical temperature:
The Green's functions $G_{ij}(E)$ for spins on impurity sites $i$ and $j$,
satisfy  
\begin{eqnarray}
EG_{ij}(E)&=&2\lambda_i\delta_{ij}+\left(\sum_l J_{lj}\lambda_l\right)G_{ij}(E)\\ \nonumber
& &-\epsilon\left[ \lambda_i\sum_lJ_{il}G_{lj}(E)\right]
\end{eqnarray}
where the variables ${\lambda_i}$ are the average magnetization on individual
sites, normalized with respect to the  magnetization averaged over all impurities.
For an RPA treatment of Heisenberg spins the  value of $\epsilon$ is 1. In order to  
compare to approximations that we 
shall term ``Ising-like'', this term can be  neglected, ie $\epsilon$ taken to be zero.
For a given temperature the Green's functions for impurity spins are determined
following a self-consistent procedure for the $G_{ij}(E)$ and ${\lambda_i(T)}$ similar to that of Callen\cite{Callen,Georges2}.
In the limit of T $\rightarrow $ T$_c$ and  a total of  $N_0$  classical spins, we can write:

\begin{eqnarray}
F_i  = - {1\over {2\pi\lambda_i}}\int^{\infty}_{-\infty}{{\rm Im} G_{ii}(E) \over {E}} dE
\end{eqnarray}
\begin{eqnarray}
k_{\rm B} {\rm{T}_c} &=& {2\over {3N_0}} \sum_{i} {1\over F_{i}}
\end{eqnarray}

T$_c$ is now determined by the self consistency of these equations,
which  are solved  exactly
for a given  configuration. The critical temperatures are 
averaged over different samples (typically 10$^5$ host sites averaged
over 50 configurations).
We remark that  
the results are close
to recent results  obtained by a Monte Carlo simulations\cite{mc}, which 
also take into account fluctuations in the positions of the 
magnetic impurities.
Semi-analytical calculations are intrinsically much faster
and essentially 
no finite-size extrapolation is needed.
For a given configuration,
equations 2 and 3 give an explicit form of T$_c$. 
The greatly improved accuracy over previous mean field methods
is due essentially to the fact that the RPA approximation
includes  the low temperature modes that destroy long range order
as the temperature increases, and that the geometric disorder is
fully included.
\begin{figure}[tbp]
\centerline{
\psfig{file=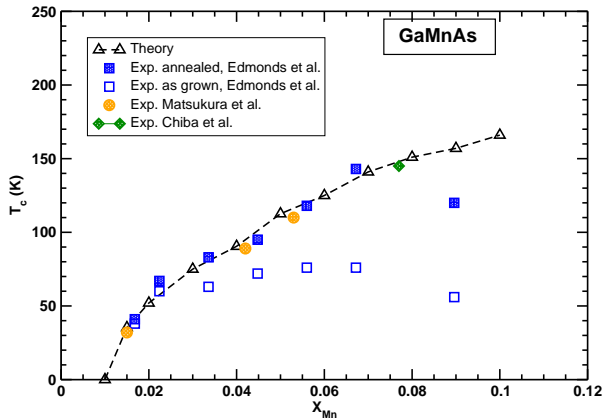,width=7cm,angle=-90}}
\caption{T$_c$ as a function of doping for Mn$_{x}$Ga$_{1-x}$As: theory for
 uncompensated samples  and 
experiment.}
\end{figure}
\par
In Figure 1 we show the calculation of T$_c$ as a function of x for 
 Mn$_{x}$Ga$_{1-x}$As.
We show also the experimental results of Edmonds et al\cite{Edmonds2,Edmonds}, Matsukura \cite{Matsukura},
and Chiba et al.\cite{Chiba} for different concentrations of 
of  Mn in  Mn$_{x}$Ga$_{1-x}$As.
We note that the agreement with the results of Edmonds et al.
is for the case of {\it fully} annealed samples. This is consistent with the fact
that we use the couplings calculated for uncompensated samples.
The annealing   changes the compensation via the density of Mn interstitials and As antisites.
This  also increases the density
of carriers, as shown 
by transport studies.
The agreement with experiment is excellent,
except for the single highest concentration (9\%):  our theory  suggests that 
at this concentration annealing  is not complete.
Note also that the theory correctly predicts a threshold ( about 1.5\%) below which there
is no ferromagnetism.  
\begin{figure}[tbp]
\centerline{
\psfig{file=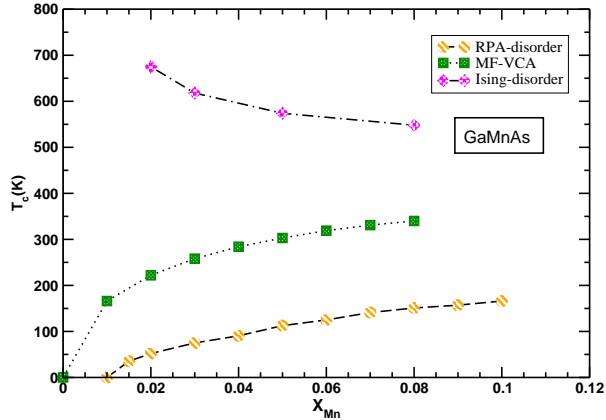,width=7cm,angle=-90}}
\caption{T$_c$  for Mn$_{x}$Ga$_{1-x}$As: comparison of different theories, as defined in the text.}
\end{figure}

In Figure 2 we show the calculations of T$_c$ as a function of x for 
 Mn$_{x}$Ga$_{1-x}$As within different theoretical approaches:
``Ising'' like Mean Field theory ($\epsilon  =0 $ in equation 1, but fully including
disorder),
, Mean Field-Virtual Crystal Approximation (MF-VCA) in which the disorder is treated as a simple effective medium:
$T^{MF-VCA}_c={2\over 3}x\sum_iJ_{0i}$, RPA-disorder ($\epsilon$ =1, full disorder and transverse fluctuations, the present approach).
One can see the large over-estimate of the critical temperatures for
the ``Ising-like'' mean field approximations (essentially the same as
in ref\cite{Bhatt} ) 
 which includes the disorder correctly but
does not treat transverse fluctuations.
In such 
simple mean field theories the critical temperature is overly influenced 
by sites in large local effective fields due to  strong short-range 
ferromagnetic interactions.
For 
low dilution long-range order cannot propagate simply by nearest neighbour
interactions. 
The RPA form, in contrast, gives more weight to the low frequency
excitations and this is the reason for its success. 
The MF-VCA results reproduce essentially 
results  of Sato et al.\cite{Satoetal},
showing that the difference with our final theory is in the treatment of 
the effective Hamiltonian, not in the  couplings estimated {\it ab initio}.
The current theory
and Monte-Carlo simulations\cite{mc} show the threshold effect for ferromagnetism: this is 
an example of the failure of the simplest RKKY-like theories.
\par
The flexibility and accuracy of our calculation allows us to 
make more precise the question of the dependence of the 
Curie  temperature on the density of carriers.  In Figure 3 we present
the estimated critical temperature as a function of the number of 
carriers for the dopant concentration 5\%. 
For direct experimental comparison, annealing may also change the number of 
magnetically active impurities through the variation of the density of  Mn interstitials.
For simplicity we analyse the effect of carrier density assuming a 
constant magnetic impurity density.
The exchange integrals are calculated by introducing compensation via a concentration $y$ of  As antisites, giving a carrier density n$_h=x-2y$.
It is seen that the temperature is rather insensitive
to hole density provided that it is above  a value of about ~ 60 \%
of the number of dopant atoms. Below this value it diminishes rapidly and
ferromagnetism disappears between 50 and 60\%. From the dependence
of the coupling constants on concentration $x$ ( see Figure 5a of ref. \cite{Josephetal}) we expect the 
threshold value of $\gamma$ to decrease further as $x$ increases. 
This transition is associated with suppression of the ferromagnetism
presumably
in favour of a spin-glass phase,
owing to  dominance  of antiferromagnetic
superexchange.
For comparison, we show the results if the disorder is treated by a Virtual
Crystal Approximation (MF-VCA). While that approximation gave a threshold
level of carrier density at rather lower values, it increasingly overestimates the 
critical temperature as $\gamma$ increases, ie the compensation decreases.

\begin{figure}[tbp]
\centerline{
\psfig{file=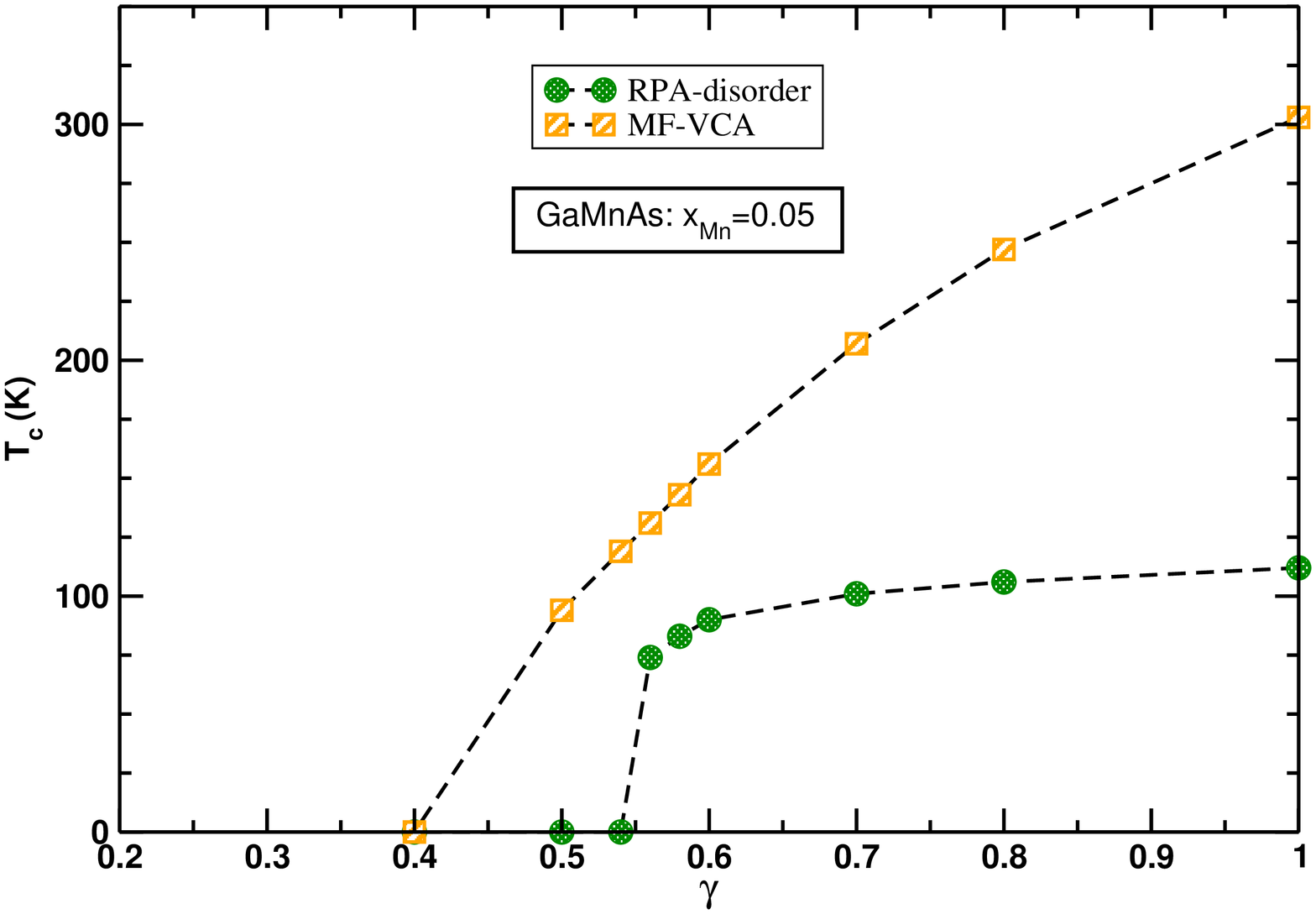,width=7cm,angle=-90}}
\caption{T$_c$  for Mn$_{x}$Ga$_{1-x}$As for fixed concentration as a function
of carrier density. The parameter $\gamma$ is the ratio
of carrier concentration n$_h$ to the doping density x.}
\end{figure}
It would be interesting to verify this curve
experimentally in different samples. 
The reference [\cite{Edmonds2} ] estimates that in as-grown samples
$\gamma$ is about $0.5$ for 5\% Mn.
Annealing increased
$T_c$ substantially: this is consistent  with
our estimates of the transition region. We note that older, possibly less precise, measurements 
of carrier concentrations\cite{Matsukura} gave lower values ($\gamma \approx 0.2 $) where we
would predict that ferromagnetism is unstable. This lower value has  been used in RKKY-type models
which failed to predict the spin-glass instability. This was  due to the 
neglect of superexchange contributions important especially at low carrier concentration\cite{BouzerarComment}.
This is  in contrast
to the {\it ab initio} calculations where they are fully included.
The relative insensitivity
with the degree of compensation indicates
that our previous calculations ( Figure 1) should remain accurate
in presence of a small number of antisite substitutions.
We remark that Figures 1 to 3  indicate the qualitative and quantitative failures
of simplified RKKY approaches in which 
T$_c \propto x^{4\over3}{\gamma}^{1\over 3}$. 
While this has already been noted experimentally\cite{Edmonds}
we can see from a theoretical point of view while this must be so:
each  exchange in the effective Hamiltonian is renormalized by the disorder,  disorder in the effective Hamiltonian  is
significant, and transverse spin fluctuations are crucial. 
\par
We now turn to the calculations made by the same procedure for the
case of (GaMn)N which  has attracted great interest\cite{DietlGaN}.
Application of simple RKKY
models predicted a T$_c$ as high as 700 K for moderately low
concentrations (6 \% Mn)\cite{DietlMatsukuraOhno}. The experimental situation is somewhat controversial:
there have been  reports of ferromagnetism at high temperatures\cite{GaNhigh}
as well as reports of paramagnetism down to low temperatures\cite{Cibert}. Here we
shall content ourselves with estimating the critical temperatures
for well-annealed samples. In Figure 4 we show predicted Curie temperatures
for Mn$_{x}$Ga$_{1-x}$N showing, for comparison,
the same calculations for Mn$_{x}$Ga$_{1-x}$As. It is seen that the 
critical temperature of the doped GaN is always lower than for the 
same concentration of doped GaAs. This is despite the fact that
the nearest neighbour ferromagnetic coupling is substantially stronger:
propagation of long range magnetic order naturally depends crucially on 
the further nearest neighbour couplings which are much weaker\cite{Josephetal}.
The reason that they fall off more rapidly is that the impurity band  of
Mn is near mid-gap, in contrast to the case of (Mn,Ga)As where
it is close to the valence band edge of GaAs.
As in GaAs, the 
estimates by Sato et al\cite{Satoetal} are qualitatively different
because of the mean field treatment, 
with maximum critical temperatures  for GaN  at
~350 K, much higher    
than the current results. We emphasise that the 
this difference is not because
of the difference in the exchange couplings, but in the
subsequent treatment of the effective Heisenberg model.
\begin{figure}[tbp]
\centerline{
\psfig{file=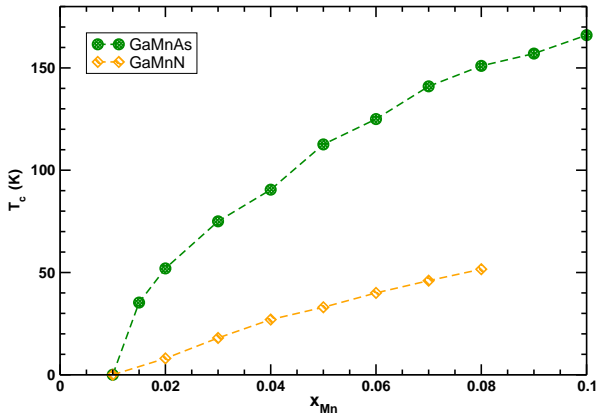,width=7cm,angle=-90}}
\caption{Predicted T$_c$ for (Ga,Mn)N compared to (Ga,Mn)As.}
\end{figure}

\par
In conclusion, we present  a novel approach allowing for reliable  calculation,
starting from {\it ab initio} methods, of the critical temperatures in 
diluted magnetic semi-conductors. This shows that apparent discrepancies
in the past between predictions from {\it ab initio} methods and
observed material properties
has been due to the incomplete treatment of the effective Hamiltonian
rather than inaccuracy in the effective exchange couplings, especially
because of the need to properly include the effects of disorder.
The speed and flexibility of the method
allows us to consider parameters that may be useful in designing useful materials in a variety of non-translationally invariant situations. Good agreement is found with experimental data on well annealed samples of 
 Mn$_x$Ga$_{1-x}$As.
The apparent strong increase of T$_c$ due to proper treatment of disorder, 
as observed in previous 
``Ising-like'' mean field theories, is shown to be due to neglect
of transverse fluctuations. These  in fact {\it reduce} the critical temperatures
significantly compared to simple MF-VCA.  
 We also predict the dependence
on the carrier concentration, corresponding to samples with 
different concentrations of anti-sites. For the average dopant concentrations chosen,
this predicts weak dependence on anti-site concentration when this is small
but eventually the critical temperature falls rapidly. 
For doped samples of GaN our theory predicts consistently {\it lower}
critical temperatures than for GaAs despite the larger
short range ferromagnetic coupling. Our estimates differ by an order 
of magnitude from previous estimates in the concentration range 0-10\%.
These results
call for explicit experimental verification and also show the over-simplification
of previous theories.

We would like to thank Dr. K. Edmonds for providing unpublished additional
data concerning measured critical temperatures of (GaMn)As.
JK acknowledges the financial support from the Grant agency of 
the Academy of Sciences of the Czech Republic (A1010203 ) and 
the Czech Science Foundation 202/04/0583).

\end{document}